\documentstyle[12pt]{article}
\newcommand{\be}{\begin{equation}}
\newcommand{\ee}{\end{equation}}
\begin{document}
{\large
\begin{center}
{\large \bf Quantum electrodynamics of composite fermions.}

Serge B. Afanas'ev

{\it Sankt--Petersburg, Russia}

e-mail:~serg@ptslab.ioffe.rssi.ru
\end{center}

\begin{abstract}
The quantum Dirac--like equation and the QED vertex operator for a
composite particle are suggested. The vertex operator and the
fermionic propagator
are connected by the QED Ward identity.
It is shown that all of the Feynman
QED--integrals for a three--component particle are finite with
some suggestion
about large momentum dependence of the
vertex operator. This dependence is
in agreement with the component counting rules.
\end{abstract}
     The ultraviolet divergences
appearance in calculations is one of the most difficult
problem of the quantum field
theory. At the present time
many methods of ultraviolet divergences elimination are developed.
Thus the ultraviolet divergences are not in the superstring theory,
and this fact is connected
with the finite sizes of superstrings as the
space--time objects. In this
work the quantum electrodynamics of a composite particle with the vertex
operator and the motion equation different from
the Dirac--Feynman--Schwinger
pointlike QED ones is considered.

Pointlike QED based on the Dirac equation
\be
({\widehat p} - m) \psi = 0
\ee

The general form of the vertex operator consequential
from requirements
of gauge and relativistic invariances is
\be
\label{vert1}
\Gamma^{\mu} (q^2) = f (q^2) \gamma^{\mu} - {1 \over {2 m}} g (q^2)
\sigma^{\mu \nu} q_{\nu},
\ee
where $q_{\mu}$ is the photon momentum.

The calculations with pointlike form factors in each exact vertex
$f (q^2) = 1$, $g (q^2) = 0$ yield the ultraviolet divergences caused
by momentum dependence of integrand
in the large momentum region.

It is noted that allowing the momentum dependence of form factors
$f(q^2)$ and $g(q^2)$ results in eliminating ultraviolet divergences~ of the
Feynman integrals
in majority
of QED--calculations~ if $f(q^2)$ and $g(q^2)$
are decreasing functions of momentum in the large momentum
region. Asymptotical momentum dependence of composite particle form factors
is described with functions $f (q^2) \sim (1/q^2)^{n-1}$, where $n$ is a
number
of components~\cite{brodsky1}.
In agreement with the components counting rules
the suggestion about leptons substructure
leads to elimination of all ultraviolet divergences, besides the polarization
operator divergence
and divergences in other Feynman diagrams in which all of external lines are
photon. Leptons substructure and the upper limits of
leptons sizes are discussed
in~\cite{lept,preon}.

However, $q^2$--dependence of the form factors leads to
the contradiction with the Ward identity:
\be
\label{ward}
G^{-1}(p+q) - G^{-1}(p) = q_{\mu} \Gamma^{\mu} (p,p+q;q)
\ee
because the one puts very rigid conditions on the form
of the fermionic propagator.
The identity (\ref{ward})
cannot be satisfied
if form factors depend on
$q^2$ only.

Suppose that the vertex operator can depend on the
lepton momentum as well as the
photon
momentum. Obviously, $p$--dependence
of the vertex operator leads to a modification of the particle propagator,
as the Ward identity (\ref{ward}) shows.

Consider the following system
consisting of the
composite particle quantum relativistic
motion equation:
\be
\label{dir}
(f(p^2) {\widehat p} - m) \psi = 0
\ee
and the vertex operator in the form
\be
\label{vert2}
\Gamma^{\mu} (p,p+q;q) = h_1(p^2, q^2) \gamma^{\mu} + h_2(p^2,q^2)
\gamma_{\nu} p^{\nu} q^{\mu}
\ee
In (\ref{vert2})
the "spin term" is omitted as it does not take a contribution
to the Ward identity due to
antisymmetric spin tensor presence.

Define the fermionic propagator
as the Green function of the equation (\ref{dir}):
\be
\label{prop}
G(p) = { {f(p^2) {\widehat p} + m} \over {f^2(p^2) p^2 - m^2}}
\ee

    Let us suppose that $f(p^2)$ is the
decreasing function of $p^2$ and
\be
\label{p=0}
f(p^2)|_{p^2 = 0} = 1
\ee
The equation (\ref{dir}) and the propagator (\ref{prop})
under the $p^2 \to 0$ conditions have
pointlike QED -- forms. If
\be
\label{finf}
f(p^2) |_{p^2 \to \infty} \sim {1 \over {p^{\alpha +1}}}, \alpha >0
\ee
then the propagator (\ref{prop})
\be
\label{prinf}
G(p) |_{p^2 \to \infty} \to - {1 \over m}
\ee

Poles of the propagator (\ref{prop}) are yielded by the equation
\be
\label{pol}
f^2(p^2) \cdot p^2 = m^2
\ee
As is shown below, the function $f(p^2)$ is identical to the
elastic form factor in the small momentum region. Therefore it
may be noted that
$|f(p^2)| << 1$ in the region
$p^2 \gg\quad<r^2>^{-1}$
where $<r^2>$ is the average composite particle size squared.
Thus if $m^2 <r^2>\quad \ll 1$ then the propagator poles are
$p^2 \simeq m^2$ and $p^2 \sim\quad<r^2>^{-1}$. In the case
$m^2 <r^2>\quad \sim 1$
both poles are in the region $p^2 \sim m^2$. If
$m^2 <r^2>\quad \gg 1$ then poles of propagator are absent.

Solutions of the equation (\ref{dir}) for a particle without external
field are not eigenstates of the momentum operator. Thus the solutions
of (\ref{dir}) are some linear combinations of the plane waves.

Considering the equation (\ref{dir})
and the vertex operator (\ref{vert2}) the Ward identity (\ref{ward}) is:
\be
f((p+q)^2) ({\widehat p} + {\widehat q}) - f(p^2) {\widehat p} =
q_{\mu} (h_1(p^2,q^2) \gamma^{\mu} + h_2(p^2,q^2) \gamma_{\nu} p^{\nu} q^{\mu})
\ee
This equation puts the following conditions on the functions
$f(p^2)$ , $h_1(p^2,q^2)$ and $h_2(p^2,q^2)$:
\be
\label{fp1}
f((p+q)^2) - f(p^2) = q^2 h_2(p^2,q^2)
\ee

\be
\label{fp2}
f((p+q)^2) = h_1(p^2,q^2)
\ee

The functions $h_1(p^2,q^2)$ and $h_2(p^2,q^2)$ are
exactly determined by the equations
(\ref{fp1}), (\ref{fp2}) and are connected by these equations.

The last equation leads to the condition
\be
h_1(p^2_1,p^2_2) = h_1(p^2_2,p^2_1)
\ee

Analyzing the Ward identity (\ref{ward}) with the vertex operator
in the form (\ref{vert2}), (\ref{fp1}), (\ref{fp2}) we come to the
conclusion that this $\Gamma^{\mu}$ is orthogonal to $q_{\mu}$.
Thus the system of the composite particle quantum
equation (\ref{dir}) and the
vertex operator (\ref{vert2}), (\ref{fp1}), (\ref{fp2}),
is gradient-invariant
and satisfies the electric charge conservation law.

The vertex operator (\ref{vert2}) is analogous to
the vertex functions
that appear at the investigation of the deep
inelastic scattering.
There can be suggested that the operator (\ref{vert2}) describes the internal
processes in a composite system. $p^2$--dependence of the function
$h_1(p^2,q^2) = f(p^2)$ even at the zero momentum transmission
describes the internal processes, but not the interaction of a composite
particle
with the electromagnetic field. This dependence coincides with
$p^2$--dependence of the function $G^{-1} (p^2)$.

Consider the ultraviolet divergences in
composite particle QED (\ref{dir}),
(\ref{vert2}). In this theory the photon propagator
has the standard QED--form.
In the large momentum region the fermionic propagator is finite. A number of
integrations over internal momenta of a diagram with $N$ vortices,
$N_f$ external
fermionic lines and $N_{ph}$ external
photon lines, is $2(N - N_f - N_{ph} + 2)$.
Thus the index of a diagram $(N,N_f,N_{ph})$ equals
\be
\label{ng}
\Omega_{N~N_f~N_{ph}} = 2 (N - N_f - N_{ph} + 2)
- 2 {{N - N_{ph}} \over 2} - N_{\Gamma} \cdot N
\ee
where $N_{\Gamma}$ is the power
of momentum in the expression for the vertex operator.
Hence
\be
\label{n}
\Omega_{N~N_f~N_{ph}} = N - 2 N_f - N_{ph} + 4 - N_{\Gamma} \cdot N
\ee
It is known that the most critical divergence is in the polarization operator
($N_{ph} = 2, N_f = 0, N = 2$). In this case (\ref{n}) yields:
\be
\Omega_{2~0~2} = 4 - 2~N_{\Gamma}
\ee
Thus if $N_{\Gamma} > 2$ then all of Feynman integrals are finite.

Consider possible large momentum dependence of the vertex operator
(\ref{vert2}), (\ref{fp1}), (\ref{fp2}). In the approximation $p^2 \to 0$
this vertex operator has the following form:
\be
\Gamma^{\mu} \to h_1(0,q^2) \gamma^{\mu}
\ee
i.e. the vertex operator (\ref{vert1}) without "spin term".
Hence we might suppose that
\be
\label{comp}
h_1(p^2,q^2) |_{{p^2 \to \infty} \atop {q^2 \to \infty}} \to {1 \over
(p+q)^{2(n-1)}}
\ee
in agreement with the components counting rules.
$h_2(p^2,q^2)$ has analogical $p^2,q^2$--dependence.
Thus with the supposition (\ref{comp}) in the case $n=3$ the Feynman
integrals are
finite for all diagrams. In the case $n=2$ ultraviolet divergence remain
in the polarization operator.

This work approach is not consistently one to the Feynman
diagram calculations
as well as the Feynman rules in pointlike QED where the vertex
operator form is consequence of the Dirac equation with the interaction
taking into account. It is to be supposed that the equation (\ref{dir})
and the vertex operator (\ref{vert2}), (\ref{fp1}), (\ref{fp2}) could be
obtained from multifermionic Green functions consideration.

The information about the momentum dependence of the form factors
could be found from concrete models of a
composite particle with knowledge
of the components interactions.
The functions $f(p^2), h_1(p^2,q^2), h_2(p^2,q^2)$ for leptons
could be obtained from preon models~\cite{preon}.
Hence this problem is not a pure
QED--problem as the electron structure cannot be determined by
electrodynamics. Thus the fact of the vertex operator and the
lepton propagator
being in principle written in the form that allows for calculating
QED--quantities without divergences has a principal character only.
The QED--calculations with the account of
small corrections, which are proportional to
$<r^n_{lept}>$, is possible.

The author thanks
T. Kinoshita for valuable questions, that enabled deeper look on
the problem,
S.J. Brodsky for the support of this work
and M. Tyntarev and M. Golod for their technical help.

\end{document}